\documentclass{PoS}

\newcommand\Ref[1]     {Ref.\,\cite{#1}}
\newcommand\Refs[1]    {Refs.\,\cite{#1}}
\newcommand\eqn[1]     {Eq.\,(\ref{#1})}

\newcommand\eqnss[2]   {Eqs.\,(\ref{#1})--(\ref{#2})}

\newcommand{\beq}      {\begin{equation}}
\newcommand{\eeq}      {\end{equation}}
\newcommand{\beeq}     {\begin{eqnarray}}
\newcommand{\eeeq}     {\end{eqnarray}}

\newcommand\as         {\alpha_{\rm s}}

\newcommand\rd         {{\rm d}}
\newcommand\nn         {\nonumber}

\newcommand\bom[1]     {{\mbox{\boldmath $#1$}}}

\newcommand{\bI}       {\bom{I}}
\newcommand{\bT}       {\bom{T}}

\newcommand\Oe[1]      {\ensuremath{\mathrm O(\eps^{#1})}}
\newcommand{\ep}       {\epsilon}        
\newcommand{\eps}      {\varepsilon}

\newcommand{\PS}[2]    {\rd\phi_{#1}^{#2}}
\newcommand\tsig[1]    {\sigma^{\mathrm{#1}}}
\newcommand\dsig[1]    {\rd\sigma^{{\rm #1}}}
\newcommand\dsiga[2]   {\rd\sigma^{{\rm #1,A}_{\scriptscriptstyle #2}}}

\newcommand{\rB}{{\rm B}}
\newcommand{\rR}{{\rm R}}
\newcommand{\rV}{{\rm V}}
\newcommand{\rA}{{\rm A}}
\newcommand{\rLO}{{\rm LO}}
\newcommand{\rNLO}{{\rm NLO}}
\newcommand{\rNNLO}{{\rm NNLO}}

\newcommand\ldot       {\!\cdot\!}
\newcommand\la         {\langle}
\newcommand\ra         {\rangle}

\newcommand\SME[3]     {|{\cal M}_{#1}^{(#2)}{(#3)}|^2}
\newcommand\M[2]       {\ensuremath{|{\cal{M}}_{#1}^{#2}|^2}}
\newcommand\bra[3]     {\la {\cal M}_{#1}^{#2}#3|}
\newcommand\ket[3]     {|{\cal M}_{#1}^{#2}#3\ra}

\newcommand{\bC}[1]    {\bom{\mathrm C}_{#1}}

\newcommand{\bS}[1]    {\bom{\mathrm S}_{#1}}

\newcommand{\bSCS}[1]  {\bom{\mathrm C}\kern-2pt\bom{\mathrm S}_{#1}}

\def\hP{\hat{P}}

\newcommand{\cSCS}[1]  {{\cal C}\kern-2pt{\cal S}_{#1}^{~}}

\title{Perturbation theory of computing QCD jet cross sections beyond NLO
accuracy}

\ShortTitle{Perturbation theory of computing QCD jet cross sections
beyond NLO accuracy}

\author{\speaker{Zolt\'an Tr\'ocs\'anyi}
\thanks{I thank the Galileo Galilei Institute for Theoretical Physics
for the hospitality during this conference.
This research was supported by the Hungarian Scientific Research Fund
grant OTKA K-60432 and by the Swiss National Science Foundation (SNF)
under contract 200020-117602.}
\\
University of Debrecen and Institute of Nuclear Research of the
Hungarian Academy of Sciences \\
H-4001 Debrecen P.O.Box 51, Hungary\\
E-mail: \email{Zoltan.Trocsanyi@cern.ch}}

\author{G\'abor Somogyi\\
University of Z\"urich \\
Winterthurerstrasse 190, CH-8057 Z\"urich, Switzerland\\
E-mail: \email{sgabi@physik.unizh.ch}}

\abstract{
We discuss the problems that arise when one wishes to extend the
existing general methods of computing radiative corrections to QCD jet
cross sections to beyond next-to-leading order. Then we present a
subtraction scheme that can be defined at any order in perturbation
theory. We give a status report of the implementation of the method to
computing jet cross sections in electron-positron annihilation at the
next-to-next-to-leading order accuracy.
         }

\FullConference{8th International Symposium on Radiative Corrections \\
		 October 1-5, 2007\\
		 Florence, Italy}

\begin{document}


Accurate predictions of QCD jet cross sections require the computation of
radiative corrections at least at next-to-leading order (NLO) accuracy,
but in some cases also at higher order. The physical cases when
computations at the next-to-next-to-leading order (NNLO) are important
have been discussed extensively in the literature
\cite{Glover:2002gz}. There are also some less standard
motivations. Although several general methods exist for computations at
NLO, these become very time-consuming with increasing number of
final-state partons. Many talks at this conference showed impressive
progress in computing one-loop corrections to amplitudes of
multileg processes \cite{OneLoopTalks}, therefore, we should consider
the question wether we can compute the real radiation corrections fast
enough. Furthermore, the thorough understanding of computing NNLO
corrections may help the combination of parton showers and NLO
calculations.  


The perturbative expansion of any jet cross section can formally be written as
$
\sigma = \sigma^{\rLO} + \sigma^{\rNLO} + \sigma^{\rNNLO}
+ \ldots\;.
$
Let us consider $e^+e^- \to m \mbox{ jet}$ production, when
$\sigma^{\rLO}$ is the integral of the fully exclusive Born cross section
over the available phase space defined by the jet function $J_m$,
\beq
\sigma^{\rLO} = \int_m \rd \tsig{B}_m J_m \equiv \int \PS{m}{}\M{m}{(0)} J_m
\,.
\eeq
There are two contributions to the NLO
correction. We have to consider the fully exclusive cross section
$\rd\sigma^\rR$ for producing $m+1$ partons and the one-loop correction
$\rd\sigma^\rV$ to the production of $m$ partons,
\beeq
\sigma^\rNLO = \int_{m+1}\rd\sigma^\rR J_{m+1} + \int_m\rd\sigma^\rV J_m
 =
\int \PS{m+1}{}\M{m+1}{(0)} J_{m+1} + \int \PS{m}{}
2\mathrm{Re} \langle {\cal M}_{m}^{(1)}|{\cal M}_{m}^{(0)}\rangle J_{m}
\,.
\eeeq
These two contributions are separately divergent in $d=4$ dimensions
although their sum is finite for infrared safe observables. We assume
that ultraviolet renormalization has been carried out, so the
divergences are purely of infrared origin and are regularized by
defining the integrals in $d=4-2\eps$ dimensions. 

There are several general methods of computing the finite NLO
correction. Most of these rely on the same principles, namely one
defines approximate cross section $\rd\sigma^\rA$ which regularizes the
real correction in $d$ dimensions in all its infrared singular limits
that lead to poles in $\ep$, so the cross section 
\beq
\sigma^{\rm NLO}_{m+1} = \int_{m+1}\!\left[
  \left(\rd\sigma^\rR\right)_{\eps=0} J_{m+1}
- \left(\rd\sigma^\rA\right)_{\eps=0} J_m
\right]
\eeq
is finite.%
\footnote{The subtraction terms in these equations are symbolic in the
sense that these are actually sums of different terms.  The jet
function depends on different momenta in each of these terms, the exact
set of momenta for each term can be found in \Ref{Somogyi:2006da}.}
The approximate cross section is constructed using the universal
soft- and collinear factorization properties of QCD matrix elements (we
use the colour-state notation \cite{Catani:1996vz} and also some notation
introduced in \Ref{Somogyi:2005xz}),%
\footnote{We drop some numerical factors in order to keep the
expressions as simple as possible, as only the structure of these
formulae is relevant for the discussion.}
\beq
\bS{r}\SME{m+1}{0}{p_r,\ldots}
\propto
\sum_{\stackrel{\scriptstyle{i,k}}{\scriptstyle{i\ne k}}}
\frac{s_{ik}}{s_{ir}s_{kr}}
\,\bra{m}{(0)}{(\ldots)} {\bom T}_{i}\ldot{\bom T}_{k} \ket{m}{(0)}{(\ldots)}
\,,
\eeq
\beq
\bC{ir}\SME{m+1}{0}{p_i, p_r,\ldots}
\propto
\frac{1}{s_{ir}}
\bra{m}{(0)}{(p_{ir},\ldots)}
\hP_{ir}^{(0)}
\ket{m}{(0)}{(p_{ir},\ldots)}
\,.
\eeq
These factorization formulae allow for such a construction that the
integration over the phase space of the unresolved parton can be computed
independently of the jet function, leading to
\beq
\int_1 \rd\sigma^\rA = \rd\sigma_m^\rB \otimes \bI(\ep)
\,,
\label{eq:intsigmaA}
\eeq
where $\bI(\ep)$ is an operator in the colour space with universal pole
part,
\beq
\bI(\ep) \propto \frac{\as}{2\pi}
\sum_i \left[
  \frac{1}{\ep} \gamma_i
- \frac{1}{\ep^2} \sum_{k\ne i} \bT_i\cdot\bT_k
\left(\frac{4 \pi \mu^2}{s_{ik}}\right)^\ep\right]
+ \Oe{0}
\,.
\eeq
This pole part is equal, but opposite in sign to the pole part of the
virtual correction, so that the $m$-parton integral
\beq
\sigma^{\rNLO}_m = 
\int_m\!\left[\rd\sigma^\rV + \int_1 \rd\sigma^\rA \right]_{\eps=0}
\label{eq:rNLOm}
\eeq
is finite. Therefore, the sum of the two finite contributions
$\sigma^{\rm NLO}_m$ and $\sigma^{\rm NLO}_{m+1}$ is equal to 
$\sigma^{\rNLO}$.

In constructing the approximate cross section special care is needed to
avoid double subtractions in the regions where a soft parton becomes also
collinear to another hard parton. At the NLO accuracy, the overlap of
the soft- and collinear limits can easily be identified to be the
collinear limit of the soft factorization formula \cite{Somogyi:2005xz}.
However, disentangling the multiple unresolved limits at higher orders,
when multiple soft-, collinear- and soft-collinear limits overlap in a
complicated way, is far more cumbersome  \cite{Somogyi:2005xz}. This calls
for a simple and systematic procedure.  

In a physical gauge the collinear singularities are due to the
collinear splitting of an external parton \cite{collinear}.
The overall colour structure of the event does not change, the
splitting is entirely described by the Altarelli--Parisi functions
which are a product of colour factors and a kinematical function
describing the collinear kinematics of the splitting. Thus, if we want
to identify the collinear contributions in the soft factorization
formulae to {\em any order in perturbation theory}, we can use the
following simple procedure: (i) employ the soft insertion rules
\cite{Bassetto:1984ik,Catani:1999ss} to obtain the usual expression
\beq
\bS{r} \SME{m+1}{0}{p_r,\dots} \propto
\sum_{i=1}^m \sum_{k=1}^m \sum_{\rm hel.}
\eps_\mu(p_r) \eps_\nu^*(p_r) \frac{2 p_i^\mu p_k^\nu}{s_{ir} s_{kr}}
\bra{m}{(0)}{(\dots)}\bT_i\ldot \bT_k\ket{m}{(0)}{(\dots)}
\,,
\eeq
with
\beq
s_{ir} = 2 p_i\ldot p_r \quad{\rm and}\quad
 \sum_{\rm hel.} \eps_\mu(p_r) \eps_\nu^*(p_r) =
-g^{\mu\nu} + \frac{p_r^\mu n^\nu + p_r^\nu n^\mu}{p_r\ldot n}
\,;
\eeq
(ii) choose Coulomb gauge ($n^\mu = Q^\mu - p_r^\mu\,Q^2/s_{rQ}$,
$s_{rQ} = 2 p_r\ldot p_Q$) to identify the collinear contribution in
the colour diagonal terms 
\beeq
&&
\bS{r} \SME{m+1}{0}{p_r,\dots} \propto
\sum_{i=1}^m \Bigg[
  \frac12 \sum_{k\ne i}^m
\Bigg(\frac{s_{ik}}{s_{ir} s_{rk}}
- \frac{2 s_{iQ}}{s_{rQ} s_{ir}}
- \frac{2 s_{kQ}}{s_{rQ} s_{kr}}\Bigg)
\bra{m}{(0)}{(\dots)}\bT_i\ldot \bT_k\ket{m}{(0)}{(\dots)}
\nn\\&&\qquad\qquad\qquad\qquad\qquad
- \bT_i^2 \frac{2}{s_{ir}} \frac{s_{iQ}}{s_{rQ}}
\SME{m}{0}{\dots}
\Bigg]
\,;
\eeeq
(iii) define momentum fractions in the Sudakov parametrization of momenta
$p_i^\mu$ and $p_r^\mu$ being collinear as
$z_i = \frac{s_{iQ}}{s_{iQ}+s_{rQ}}$, so that the colour-diagonal terms
become equal to the collinear limit of the soft factorization formula.
Then the pure soft contributions are given by
\beq
\bS{r}^{\rm pure} \SME{m+1}{0}{p_r,\dots} \propto
\sum_{i=1}^m \Bigg[
  \frac12 \sum_{k\ne i}^m
\Bigg(\frac{s_{ik}}{s_{ir} s_{rk}}
- \frac{2 s_{iQ}}{s_{rQ} s_{ir}}
- \frac{2 s_{kQ}}{s_{rQ} s_{kr}}\Bigg)
\bra{m}{(0)}{(\dots)}\bT_i\ldot \bT_k\ket{m}{(0)}{(\dots)}
\Bigg]
\,.
\eeq
We checked explicitly that this procedure leads to non-overlapping
factorization formulae that describe the analytic behaviour of the
squared matrix elements in any IR limit at the NNLO accuracy
\cite{Nagy:2007mn}. Furthermore, the factorization formula in the purely
soft limit is independent of the helicity of the soft gluon. This
allows for the definition of approximate cross sections for real
radiation with fixed helicities, and thus for Monte Carlo summation over
the helicities in NLO computations. The Monte Carlo summation over
the helicities in computations at LO was found very useful for saving CPU
time \cite{Draggiotis:1998gr}.


The physical motivation for higher accuracy and the success of the
subtraction schemes at NLO lead one to consider the extension of the
subtraction method to NNLO, when three terms contribute: the
double-real, the real-virtual and the double-virtual cross sections,
\beq
\sigma^{\mathrm{NNLO}} = \tsig{RR}_{m+2} + \tsig{RV}_{m+1} +
\tsig{VV}_{m} \equiv
  \int_{m+2} \dsig{RR}_{m+2} J_{m+2}
+ \int_{m+1} \dsig{RV}_{m+1} J_{m+1}
+ \int_{m} \dsig{VV}_{m} J_{m}
\,.
\eeq
The necessary ingredients for constructing approximate cross sections,
namely (i) the tree level three-parton splitting functions
\cite{3split}
and double soft $gg$ and $q\bar{q}$ currents
\cite{Catani:1999ss,Berends:1988zn}
and (ii) the one-loop two-parton splitting functions
\cite{loopsplitting}
and soft-gluon current \cite{Catani:2000pi} that is, the infrared (IR)
structures of the three contributions at NNLO have been known for
some time. The difficulty of using the multiple infrared factorization
formulae for cunstructing the approximate cross sections is amply
demonstrated by the slow progress in setting up a subtraction scheme.
Other approaches to computing NNLO corrections have been more
successful.  

The antennae subtraction method \cite{GehrmannDeRidder:2005cm} uses
complete squared matrix elements instead of the IR structure and the
first complete computation of NNLO corrections to three-jet production in
electron-positron annihilation has been reported to this conference
\cite{TGatRADCOR}. For processes involving massive particles and/or
simple kinematics, direct numerical evaluation of the coefficients in
the Laurent expansion of the three contributions (based on sector
decomposition) lead to the complete NNLO corrections of Higgs-
\cite{HiggsNNLO}
and vector-boson production in hadron collisions \cite{DYNNLO}

The reorganization of the NNLO contributions into three finite cross
sections,
\beq
\sigma^{\rNNLO} = \tsig{\rNNLO}_{m+2} + \tsig{\rNNLO}_{m+1} +
\tsig{\rNNLO}_{m}
\,,
\eeq
is governed by the jet function as follows:
\beq
\tsig{\rNNLO}_{m+2} =
 \int_{m+2} \Big\{ \dsig{RR}_{m+2} J_{m+2}
-\dsiga{RR}{2}_{m+2} J_m
-\Big( {\dsiga{RR}{1}_{m+2}} J_{m+1}
-\dsiga{RR}{12}_{m+2} J_m\Big)\Big\}
\,,
\label{eq:rNNLOm+2}
\eeq
\beq
\tsig{\rNNLO}_{m+1} =
 \int_{m+1} \Big\{\Big( \dsig{RV}_{m+1}
+\int_1 \dsiga{RR}{1}_{m+2}\Big) J_{m+1}
-\Big[\dsiga{RV}{1}_{m+1}
+\Big(\int_1 \dsiga{RR}{1}_{m+2}\Big){}^{\mathrm{A}_1}\Big] J_{m} \Big\}
\label{eq:rNNLOm+1}
\eeq
and
\beq
\tsig{\rNNLO}_{m} =
 \int_m \Big\{ \dsig{VV}_{m}
+\int_2\Big( \dsiga{RR}{2}_{m+2}
-\dsiga{RR}{12}_{m+2}\Big)
+\int_1\Big[\dsiga{RV}{1}_{m+1}
+\Big(\int_1 \dsiga{RR}{1}_{m+2}\Big){}^{\mathrm{A}_1}\Big]\Big\} J_m
\,.
\label{eq:rNNLOm}
\eeq

Let us concentrate on \eqn{eq:rNNLOm+1}. The construction of
$\dsig{RV,A_1}_{m+1}$ that regularizes the kinematical singularities of
$\dsig{RV}_{m+1}$ in the singly-unresolved regions is straightforward,
but the difference
$\left[\dsig{RV}_{m+1} J_{m+1} - \dsig{RV,A_1}_{m+1}  J_m\right]_{\ep = 0}$
is infinite. To make it finite, we have to subtract the universal pole
part, given by \eqnss{eq:intsigmaA}{eq:rNLOm}, too. The latter however,
does not obey universal collinear factorization. Due to coherent
soft-gluon emission from unresolved partons only the sum
$\bra{m+1}{(0)}{}(\bT_j\ldot \bT_k+\bT_r\ldot \bT_k)\ket{m+1}{(0)}{}$
factorizes in the collinear limit ($\bT_{jr} = \bT_j + \bT_r$),
\beq
\bC{jr}
\bra{m+1}{(0)}{} (\bT_j\ldot \bT_k + \bT_r\ldot \bT_k) \ket{m+1}{(0)}{}
\propto
\frac{1}{s_{jr}}
\,\bra{m}{(0)}{} \bT_{jr}\ldot \bT_k \,\hP^{(0)}_{jr} \ket{m}{(0)}{}
\,.
\eeq
This factorization is violated by the factors $s_{ik}^{-\ep}/\ep^2$ at
$\Oe{0}$, which was also noticed in \Ref{GehrmannDeRidder:2007jk}, where
it was shown that the terms that violate factorization are known to
give vanishing contribution after integration.  However, if one insists
on defining {\em fully local subtractions}, which is important for
numerical stability and reducing CPU time, then the use of properly
defined new approximate cross sections is necessary. The complete
subtraction scheme at NNLO, based on these new approximate cross sections
is defined in \Refs{Somogyi:2006da,Somogyi:2006cz}.
We employed this subtraction scheme for computing the finite cross sections
$\tsig{\rNNLO}_{m+2}$ and $\tsig{\rNNLO}_{m+1}$ of the C-parameter and
thrust distributions in electron-positron annihilation. In order to have
the complete physical prediction we also have to compute
$\tsig{\rNNLO}_m$, which requires the integration of the subtraction
terms over the singly- and doubly-unresolved factorized phase spaces.
We used standard techniques 
fractioning, sector decomposition \cite{sectordecomposition} and
residuum subtraction to find the Laurent expansion of the one-particle
integrals in
\beq
\int_1\Big[\dsiga{RV}{1}_{m+1}
+\Big(\int_1 \dsiga{RR}{1}_{m+2}\Big){}^{\mathrm{A}_1}\Big]
\,.
\eeq
We expect that the same techniques can also be employed for the
computation of the coefficients in the $\ep$-expansion of the
two-particle integral
\beq
\int_2\Big( \dsiga{RR}{2}_{m+2} -\dsiga{RR}{12}_{m+2}\Big)
\,.
\eeq
This work is in progress.

\end{document}